\documentclass[12pt]{article}
\usepackage{amsmath}
\usepackage{amsthm}
\usepackage{amssymb}
\usepackage{amsfonts}
\usepackage{latexsym}
\usepackage{url}
\usepackage{graphicx}
\usepackage{float}
\usepackage{verbatim}
\usepackage{bbm}
\usepackage{fullpage}
\usepackage{color,ctable}
\usepackage{setspace}
\usepackage[round,sort]{natbib}
\usepackage{multicol}

\newtheorem{Algorithm}{Algorithm}

\renewcommand{\epsilon}{\varepsilon}

\allowdisplaybreaks

\begin{document}

\doublespacing

    \title{Resistant Multiple Sparse Canonical Correlation}

    \author{Jacob Coleman, Joseph Replogle, Gabriel Chandler, and Johanna Hardin}

\maketitle

\abstract{Canonical Correlation Analysis (CCA) is a multivariate technique that takes two datasets and forms the most highly correlated possible pairs of linear combinations between them.  Each subsequent pair of linear combinations is orthogonal to the preceding pair, meaning that new information is gleaned from each pair.  By looking at the magnitude of coefficient values, we can find out which variables can be grouped together, thus better understanding multiple interactions that are otherwise difficult to compute or grasp intuitively.

CCA appears to have quite powerful applications to high throughput data, as we can use it to discover, for example, relationships between gene expression and gene copy number variation. One of the biggest problems of CCA is that the number of variables (often upwards of 10,000) makes biological interpretation of linear combinations nearly impossible. To limit variable output, we have employed a method known as Sparse Canonical Correlation Analysis (SCCA), while adding estimation which is resistant to extreme observations or other types of deviant data. In this paper, we have demonstrated the success of resistant estimation in variable selection using SCCA.  Additionally, we have used SCCA to find {\em multiple} canonical pairs for extended knowledge about the datasets at hand.  Again, using resistant estimators provided more accurate estimates than standard estimators in the multiple canonical correlation setting.

R code is available and documented at \url{https://github.com/hardin47/rmscca}.
}


\renewcommand{\baselinestretch}{1.75} \large \normalsize


\section{Introduction}

High-throughput data is infamous for having myriad complicated functional relationships both within and across different types of measurements on the same samples.  Multivariate statistics have been useful to understand molecular relationships by applying and modifying such techniques as principal component analysis \citep{Pea01,Zou06} and partial least squares \citep{Wol73,Ngu01}. Canonical correlation is another multivariate statistical technique used to relate two datasets evaluated on the same samples.  Recent work includes \citet{Wan14} and \citet{Hon13}, who use sparse canonical correlation to infer gene networks as tightly connected groups. \citet{LeC09} use sparse canonical correlation to compare two different microarray platforms. \citet{Gao14} provides theoretical justification of the use of sparse canonical correlation in practice.

Canonical Correlation Analysis (CCA), proposed by \citet{Hotelling}, is a multivariate method for finding linear combinations of variables in high dimensions. Given two data sets, (traditional) CCA produces as many pairs of linear combinations - called canonical pairs - as variables in the smaller set. Each canonical pair has an associated correlation, called canonical correlation, and is orthogonal to every other pair. The canonical pairs, derived through singular value decomposition of the joint covariance matrix, are ordered by their associated canonical correlations.  The goal of CCA is to maximize the canonical correlations.

While CCA is extremely useful for efficiently discerning relationships between variables, there are some drawbacks. Sensitivity to noise and outliers is one problem of CCA, and resistant CCA has only had minimal exploration in the literature \citep{Branco,Karnel91}. Especially in high dimensions, even a small amount of noise or outlying values can lead to falsely high correlations and incorrectly associated variables. To address this, we use Spearman correlations to create both correlation and covariance measures.  Through simulation, we demonstrate the need for and success of using a Spearman-like covariance estimate during CCA.  While resistant CCA might find pairs of the most highly correlated linear combinations, variable selection is somewhat limited because the output includes coefficients for {\em every} variable in both datasets.

Particularly if the number of variables is quite large, if the goal is to find highly correlated groups of variables, CCA becomes impractical.  That is, a linear combination of thousands of variables is difficult to interpret, and the analyst will be unable to discern which variables are most important. To handle the large number of coefficients reported from CCA, we employ a technique known as Sparse Canonical Correlation Analysis (SCCA) which sets some of the coefficients to zero.  \citet{Parkho} introduce SCCA and provide an algorithm for computing sparse variables, and subsequently demonstrate the success of SCCA for variable selection with a latent variable simulation model. \citet{Parkho} also demonstrate that as sample size decreases, SCCA outperforms CCA. Using a similar technique but from the perspective of Penalized Matrix Decomposition, \citet{Witten} also explore SCCA and provide the framework for computing sparse variables with different penalty functions.  In investigations of extensions of SCCA, \citet{CandF} explore different penalty functions and their relative successes on simulated data.  We use an algorithm similar to \citet{Parkho} with two modifications to first create a resistant measure and then subsequently to extend the method to find multiple canonical pairs.

As with (traditional) CCA, when using SCCA to analyze two different types of data (e.g., phenotypic and genotypic), there is typically interest in not only the first canonical relationship, but also in secondary relationships.  Using related techniques to Principal Component Analysis (PCA) where the observations are transformed into multiple linear relationships, CCA also partitions the data into linear subspaces where multiple pairs of linear relationships describe the existing patterns.  We use singular value decomposition on the cross covariance matrix to find sequential canonical pairs which are highly correlated.  Note that \citet{WittenSparse} briefly mention one idea for extending SCCA to MSCCA, but they do not assess the method or give the reader a sense of how to find the number of canonical pairs which should be considered significant.  None of the other references using or extending SCCA consider the case of more than one canonical pair.

In section \ref{SectionTheory}, we present the background mathematics of CCA \citep{Hotelling}, SCCA \citep{Parkho,Witten,CandF}, and our derivation of multiple SCCA (MSCCA) and resistant multiple SCCA (RMSCCA).  We then present our results in a series of simulations.  In subsection \ref{SectionCutoff}, we describe our process for establishing a cutoff for determining which canonical pairs are significant.  Our results comparing MSCCA and RMSCCA are given in subsection \ref{SubsectionResults}.  RMSCCA is applied to publicly available data in subsection \ref{SubsectionData}.  We conclude our work in section \ref{SectionConclusion}.

\section{Mathematical Derivation of RMSCCA \label{SectionTheory}}

\subsection{Derivation of Canonical Correlation Analysis (CCA)}

Canonical Correlation Analysis (CCA) derives pairs of linear combinations between two distinct datasets that are as highly correlated as possible.  The focus of CCA is to reveal relationships both within one group of variables and between the two sets of variables; coefficient values in one linear combination explain relationships within one dataset, while the pair of linear combinations explains relationships between datasets. Though canonical correlation does not distinguish between the explanatory and response variables, CCA can be considered in the context of multivariate regression. First developed by \citet{Hotelling}, CCA is a powerful tool for quickly determining relationships between a large number of variables.  The output of CCA will be pairs of linear combinations ordered by correlation between linear combinations, such that each linear combination is orthogonal to every preceding linear combination. The coefficients for the linear combinations are called \textit{canonical vectors}, while the linear combinations themselves are called \textit{canonical variables}. The correlations between the \textit{canonical variables} are called \textit{canonical correlations}.

Consider a pair of datasets,  ${\bf x}_{n \times p}$ and  ${\bf y}_{n\times q}$ where the columns are variables and the rows are observations, CCA finds linear combinations of the $p-$dimensional random vector {\bf X} and the $q-$dimensional random vector {\bf Y}.  The canonical vectors $\boldsymbol{\alpha}$ and $\boldsymbol{\beta}$ maximize
\begin{eqnarray*}
\rho (\boldsymbol{\alpha}'{\bf X},\boldsymbol{\beta}'{\bf Y})  &=&\frac{\boldsymbol{\alpha'\Sigma_{XY}\beta}} {\sqrt{\boldsymbol{\alpha'\Sigma_{XX}\alpha}
\boldsymbol{\beta'\Sigma_{YY} \beta}}},
\end{eqnarray*}
where
\begin{eqnarray} \label{covMatrix}
Cov({\bf X},{\bf Y}) = \begin{pmatrix}
                \boldsymbol{\Sigma_{XX}} & \boldsymbol{\Sigma_{XY}}\\
                \boldsymbol{\Sigma_{YX}} & \boldsymbol{\Sigma_{YY}}\\
                \end{pmatrix}
\end{eqnarray}

Because the scaling of $\boldsymbol{\alpha}$ and $\boldsymbol{\beta}$ does not affect the maximum correlation, CCA returns canonical vectors subject to the additional constraint:
\begin{equation}
\max_{\boldsymbol{ \alpha, \beta}} \frac{\boldsymbol{ \alpha'\Sigma_{XY} \beta}} {\sqrt{\boldsymbol{ \alpha'\Sigma_{XX}\alpha \beta'\Sigma_{YY} \beta}}} \text{ subject to }\boldsymbol{\alpha'\Sigma_{XX} \alpha} = \boldsymbol{\beta'\Sigma_{YY}\beta} = 1. \label{eq:ccamax}
\end{equation}

Once the first linear combinations are found (called the \textit{first canonical pair}), CCA maximizes the correlation between pairs of linear combinations of ${\bf X}$ and ${\bf Y}$ under the constraint that the second pair of linear combinations is orthogonal to the first. The process is repeated $\min(p,q)$ times.

The canonical correlation algorithm can be reduced to a singular value decomposition problem where $\boldsymbol{\alpha}$ and $\boldsymbol{\beta}$ are the right and left singular vectors of
\begin{eqnarray} \label{matrixK}
K = \boldsymbol{\Sigma_{XX}^{-1/2} \Sigma_{XY} \Sigma_{YY}^{-1/2}} = UDV^T
\end{eqnarray}
with $U=({\bf u}_1, {\bf u}_2, \ldots, {\bf u}_k)$ and $V=({\bf v}_1, {\bf v}_2, \ldots, {\bf v}_k)$ and $k$ is the rank of the matrix $K$.  When using real data, both $\boldsymbol{\Sigma_{XX}}$ and $\boldsymbol{\Sigma_{YY}}$ are estimated using the diagonal of the sample covariance matrices, as done by \citet{Parkho,CandF,WittenSparse}.  The $i^{th}$ canonical pair can then be represented by the $i^{th}$ singular vectors, where the canonical vectors are given by
\begin{eqnarray}
\boldsymbol{\alpha}_i &=& \boldsymbol{\Sigma_{XX}}^{-1/2} {\bf u}_i, \label{acoef}\\
\boldsymbol{\beta}_i &=& \boldsymbol{\Sigma_{YY}}^{-1/2} {\bf v}_i. \label{bcoef}
\end{eqnarray}

\subsection{Derivation of SCCA}

As with CCA, Sparse Canonical Correlation (SCCA) is a method used to create canonical vectors which represent linear combinations of two distinct datasets.  Additionally, SCCA is also based on singular value decomposition (SVD) of the covariance matrix.  Because SVD can be thought of as an iterative algorithm to find the singular vectors which lead to the decomposition, \citet{Parkho} use a SVD-like algorithm with an additional thresholding parameter to control the number of variables included in the solution of the canonical vector.  The thresholding is a form of $L_1-$regularization similar to LASSO \citep{Tib96} which sets small values of the coefficients to zero.   The algorithm due to \citet{Parkho} is for the {\em first canonical pair} and is given as follows.

\begin{Algorithm}\label{AlgorithmSCCA}
		Let $\lambda_u$ and $\lambda_v$ be chosen.  Select initial values of $\boldsymbol{u}^0$ and $\boldsymbol{v}^0$.  Set $i=0$ and $K$ to be as given in equation (\ref{matrixK}) ($i$ indexes the canonical pairs).
		\begin{enumerate}
			\item Update $\boldsymbol{u}$:
            \begin{enumerate}
                \item $\boldsymbol{u}^{i+1} \leftarrow K \boldsymbol{v}^i$
                \item Normalize: $\boldsymbol{u}^{i+1} \leftarrow \frac{\boldsymbol{u}^{i+1}}{||\boldsymbol{u}^{i+1}||}$
                \item Soft-thresholding for sparse solution:\\
                    $\boldsymbol{u}^{i+1}_j \leftarrow (|\boldsymbol{u}^{i+1}_j| - \frac{1}{2}\lambda_u)_+ Sign(\boldsymbol{u}^{i+1}_j) \mbox{ for } j = 1, \ldots, p$
                \item Normalize: $\boldsymbol{u}^{i+1} \leftarrow \frac{\boldsymbol{u}^{i+1}}{||\boldsymbol{u}^{i+1}||}$
            \end{enumerate}
			\item Update $\boldsymbol{v}$:
            \begin{enumerate}
                \item $\boldsymbol{v}^{i+1} \leftarrow K \boldsymbol{u}^{i+1}$
                \item Normalize: $\boldsymbol{v}^{i+1} \leftarrow \frac{\boldsymbol{v}^{i+1}}{||\boldsymbol{v}^{i+1}||}$
                \item Soft-thresholding for sparse solution:\\
                    $\boldsymbol{v}^{i+1}_j \leftarrow (|\boldsymbol{v}^{i+1}_j| - \frac{1}{2}\lambda_v)_+ Sign(\boldsymbol{v}^{i+1}_j) \mbox{ for } j = 1, \ldots, p$
                \item Normalize: $\boldsymbol{v}^{i+1} \leftarrow \frac{\boldsymbol{v}^{i+1}}{||\boldsymbol{v}^{i+1}||}$
            \end{enumerate}
            \item $i \leftarrow i+1$
            \item Repeat steps 1-3 until convergence.
            \end{enumerate}
where $(x)_+$ is equal to $x$ if $x \geq 0$ and 0 if $x < 0$ and
\begin{eqnarray*}
            Sign(x) = \begin{cases}
    -1 & \quad \text{if } x<0\\
    1 & \quad \text{if } x >0\\
    0 & \quad \text{if } x=0.
\end{cases}
\end{eqnarray*}
Also, define the norm of a vector $\boldsymbol{y}$ as $||\boldsymbol{y}|| = \sqrt{\boldsymbol{y}^T \boldsymbol{y}}$.
	\end{Algorithm}
	
We follow the convention of \citet{Parkho} to both set the initial canonical coefficient vectors ($\boldsymbol{u}^0$ and $\boldsymbol{v}^0$) to be the row means and column means, respectively, of the $K$ matrix and to use cross-validation to find optimal values of $\lambda_u$ and $\lambda_v$.   Because our work concerns finding multiple canonical pairs, our cross validation scheme is derived in the next section on MSCCA.  Note that $\boldsymbol{u}$ and $\boldsymbol{v}$ are sparse, and in an actual data analysis, we use diagonal versions of $\Sigma_{XX}$ and $\Sigma_{YY}$.  Therefore, the canonical vectors $\boldsymbol{\alpha}$ and $\boldsymbol{\beta}$, represented by equations (\ref{acoef}) and (\ref{bcoef}), are also sparse.

\subsection{Derivation of MSCCA}

 In CCA, sequential canonical coefficient vectors are found to simultaneously maximize the relevant correlation while maintaining orthogonality with the previous canonical coefficient vectors.  With sparse vectors, when maximizing the canonical correlation, it is necessary to choose between orthogonality and sparsity.  In order to address the sparse / orthogonality tradeoff, the singular value decomposition can be adapted to accommodate the information reduction after the first canonical pair is found.  Recall that CCA is based on SVD of the scaled cross-covariance matrix, as in equation (\ref{matrixK}).  It can be shown that the matrix $K$ can be further decomposed into singular vectors and variables.

\begin{eqnarray*}
K &=& UDV^T\\
&=& d_1 \boldsymbol{u}_1 \boldsymbol{v}_1^T + d_2 \boldsymbol{u}_2 \boldsymbol{v}_2^T + \cdots d_k \boldsymbol{u}_k \boldsymbol{v}_k^T\\
\end{eqnarray*}
where $d_i$ is the $i^{th}$ singular value.  Note that because the canonical vectors are orthogonal, the first singular value can be written as a function of the singular vectors and the matrix $K$.
\begin{eqnarray*}
\boldsymbol{u}_1^T K \boldsymbol{v}_1 = d_1.
\end{eqnarray*}
Using the ideas above for SVD, we extend the result to get the following recursive relationship:
\begin{eqnarray*}
K_{i+1} = K_{i} - (\boldsymbol{u}_{i}^T K_{i} \boldsymbol{v}_{i}) \boldsymbol{u}_{i} \boldsymbol{v}_{i}^T
\end{eqnarray*}
Each computation of the $i^{th}$ canonical pair will be based on  using $K_i$ in Algorithm \ref{AlgorithmSCCA}.

\citet{WittenSparse} mention extending SCCA to MSCCA, and use a similar derivation to the one we have provided above.  However, their SCCA algorithm is slightly different, and they provide no guidance for how to choose the number of significant pairs of canonical relationships.

Important to Algorithm \ref{AlgorithmSCCA} is the choice of $\lambda_u$ and $\lambda_v$.  The thresholding values should be optimal for a given dataset but should not overfit the data.  Note that the values of $\lambda_u$ and $\lambda_v$ for the first canonical pair will impact the decomposition of $K$ for the next canonical pair (and for all following canonical pairs).  The goal of Algorithm \ref{AlgorithmCV} is to find the optimal values of $\lambda_u$ and $\lambda_v$ for each canonical pair.

\begin{Algorithm}\label{AlgorithmCV}
		Let ${\bf x}_{n \times p}$ and ${\bf y}_{n \times q}$ represent two distinct datasets.  Set $i=1$.  Split the data into $n.cv$ cross validation partitions.  This creates $n.cv$ test sets, where the training set consists of all data not included in the particular partition. Let $K_1 = \boldsymbol{\Sigma_{XX}^{-1/2} \Sigma_{XY} \Sigma_{YY}^{-1/2}}$.   Again, when using real data, both $\boldsymbol{\Sigma_{XX}}$ and $\boldsymbol{\Sigma_{YY}}$ are estimated using the diagonal of the sample covariance matrices, as done by \cite{Parkho,CandF,WittenSparse}.
 		\begin{enumerate}
			\item Compute $\lambda_u$ and $\lambda_v$ for $i^{th}$ canonical pair:
            \begin{enumerate}
                \item Let $\lambda_u$ and $\lambda_v$ range separately along a grid of points in an interval $I_\lambda \in [0,2]$.
                \item For each ($\lambda_u$, $\lambda_v$) pair, use Algorithm \ref{AlgorithmSCCA} to find the canonical vectors and related canonical correlations (denoted $cc_{test}$) on the {\em test} data.
                \item Repeat step 1 (b) for each of the $n.cv$ choices for the test data.
                \item Choose as ($\lambda_u^*$, $\lambda_v^*$) the pair of thresholding variables that maximize the average canonical correlation ($\overline{cc}_{test}$) of the training canonical vectors applied to the test data (averaged over the $n.cv$ test data partitions).
                \item Using ($\lambda_u^*$, $\lambda_v^*$) and Algorithm \ref{AlgorithmSCCA}, find the canonical vectors based on the {\em entire} dataset to find $\boldsymbol{u}^*$ and $\boldsymbol{v}^*$.
            \end{enumerate}
			\item Adjust $K$:
                \begin{equation*}K_{i+1} = K_{i} - ({\boldsymbol{u}_{i}^*}^T K_{i} \boldsymbol{v}_{i}^*) \boldsymbol{u}_{i}^* {\boldsymbol{v}_{i}^*}^T \end{equation*}
            \item $i \leftarrow i+1$
            \item Let $pq^*$ be the number of desired canonical pairs. Repeat Steps 1-3 for $pq^* \leq \min(p,q)$ canonical pair values.
            \item Output consists of
                \begin{enumerate}
                \item A $pq^* \times 2$ matrix of ($\lambda_u^*$, $\lambda_v^*$) pairs which have maximized correlations based on training data.
                \item A pair of canonical correlations for each of the $pq^*$ canonical pairs: $cc$ on the full data set, and $\overline{cc}_{test}$, the average over all test sets for ($\lambda_u^*$, $\lambda_v^*$).

                \end{enumerate}
            \end{enumerate}
	\end{Algorithm}

\subsection{Resistant MSCCA}

In previous work on SCCA, estimation of the covariance matrix (see Equation (\ref{covMatrix})) has been done using maximum likelihood estimation.  Using maximum likelihood estimation is akin to maximizing the Pearson correlation in finding thresholding variables and canonical variables.  There has been some work in the literature on robust CCA, but, for example, \citet{Branco} considers only situations with $p$ and $q$ as large as 4; \citet{Dehon} considers only $p$ and $q$ as large as 3.

Because biological high-throughput data (and other types of data in high dimensions to which canonical correlation and its variants are often applied) are notoriously noisy, we give results on a resistant version of MSCCA applied to both multivariate normal data as well as heavy tailed data.  The methods and algorithms above are as given in the preceding algorithms except that the covariance matrices (Equation (\ref{covMatrix})) are calculated based on the ranked data as given in the {\ttfamily cov($\cdot$, $\cdot$, method="spearman")} function in R \citep{R}.  Due to the computational complexity of the algorithms, we have used a simple resistance measure.  If the user has an ability to parallelize the complete application of the algorithm, it would be worth considering other estimates of covariance like the minimum covariance determinant \citep{Rousseeuw1984}, projection pursuit \citep{Huber:1985}, or M-estimates \citep{Hardin:2007}.

\subsection{Simulating Significance Cutoff \label{SectionCutoff}}

An important step in using multiple canonical correlation pairs is deciding the number of canonical pairs to consider as significant.  In order to address concerns about multiple comparisons, we use a permutation scheme (100 permutations in the simulations below) that provides a correlation cutoff which controls the overall level of significance.  Because the process for finding canonical coefficients optimizes the respective correlation, the first canonical correlation value tends to be quite high.  Similarly, the subsequent correlations are typically decreasing but are often higher than standard correlations on most datasets.  Therefore, it is important to have a method which evaluates which canonical pairs are significant while controlling for familywise error rate.

For analyzing an actual (or simulated) dataset, we wrote the following algorithm.
\begin{Algorithm}\label{AlgorithmCutoff}
		Let ${\bf x}_{n \times p}$ and ${\bf y}_{n \times q}$ represent two distinct datasets.  Set $i=1$, ($i$ indexes the canonical pair).  Let $n.perm$ be the number of permutations.
 		\begin{enumerate}
			\item Canonical correlation values on permuted data:
            \begin{enumerate}
                \item Permute the rows of (WLOG) ${\bf y}$.
                \item Apply Algorithms \ref{AlgorithmSCCA} and \ref{AlgorithmCV} to the permuted data to find the $pq^* \leq \min(p,q)$ canonical correlations.
                \item Repeat steps 1(a) and 1(b) for $n.perm$ permutations of the original data.
                \item Let $\overline{cc}_{perm,i,(Q)}$ be the $Q^{th}$ percentile (averaged test data) correlation (across $n.perm$ correlations) for the $i^{th}$ canonical pair.
            \end{enumerate}
            \item Apply Algorithms \ref{AlgorithmSCCA} and \ref{AlgorithmCV} to the original data to find the $pq^*$ canonical correlations, $cc_{i}$ for the $i^{th}$ canonical pair.
            \item Finding significant correlations:
            \begin{enumerate}
                \item Let $j^*$ be the largest value of $j$ such that:
                    \begin{equation*}
                        \overline{cc}_{j,test} > \overline{cc}_{perm,j,(Q)} \ \ \ \forall j \leq j^*
                    \end{equation*}
            \end{enumerate}
            \item Report the canonical variables and respective canonical correlations on the original data from 1 to $j^*$.
            \end{enumerate}
	\end{Algorithm}

The algorithm allows us to report the top $j^*$ canonical pairs as significant.  Because all $j^*$ correlations are above the pointwise Q quantile of the permutation scheme, we have controlled our familywise error rate at $100-(Q)\%$ (see subsection \ref{SubsubsectionTypeI}).

The reason that the comparisons for establishing statistical significance is based on the average correlations over cross validated test sets  ($\overline{cc}_{test})$  is due to issues regarding the curse of dimensionality. We are considering cases where potentially $n<<\min(p,q)$.  Even with shrinkage induced by the penalization scheme, very high correlations are likely to be found when there is no relation between the two data sets.  By forcing the canonical vectors to act on data they were not trained on, we avoid the overfitting common with small $n$, large $p$ situations.  Only if the canonical vectors are picking up on actual signal are we then likely to see a similarly high canonical correlation on the test data.  Without this modification, we have found that it is nearly impossible to distinguish signal from noise.

\section{Simulations}
\subsection{Simulation Set-up \label{SubsectionSetup}}

In order to assess resistant multiple sparse canonical correlation (RMSCCA), we set up simulations with and without heavy tails (representing realistic noisy data). For each simulation of sample size $n$, we generate one dataset (${\bf x}$) using a multivariate normal.  Then a second dataset (${\bf y}$) is generated as a multivariate normal distribution around a linear combination of ${\bf x}$.  Similar to \citet{CandF}, we let ${\bf X} \sim MVN_p(0, \Sigma_{XX})$.  Then for each individual $l$, ${\bf Y}_l  \sim MVN_q(\mu_l, \Sigma_{YY})$, where $\boldsymbol{\mu}_l = {\bf X}_l \times B$.

The matrix ${\bf B}$ determines the relationship between ${\bf X}$ and ${\bf Y}$ and is all zeros except in coordinates to prescribe a particular relationship.  For our purposes, ${\bf B}$ is given by the equation (\ref{Bmatrix}).  Note that $1_{n\times m}$ is an $n \times m$ matrix of 1s.  Similarly, $0_{n\times m}$ is an $n \times m$ matrix of 0s.  The ${\bf B}$ matrix allows for multivariate linear relationships between ${\bf X}$ and ${\bf Y}$.  The population setup gives five sets of canonical pairs.  The first canonical pair is given by the relationship between first 10 dimensions of the random variable ${\bf X}$ and the first 20 dimensions of the random variable ${\bf Y}$; the second canonical pair is represented by the next 5 dimensions of ${\bf X}$ and the next 5 dimensions of ${\bf Y}$ variables; and so on.

				\begin{equation}\label{Bmatrix}
					{\bf B}_{p\times q} =
					\begin{pmatrix}
						1_{10\times20} & 0_{10\times5} & 0_{10\times10} & 0_{10\times50} & 0_{10\times15} & 0_{10\times q-100}\\[3pt]
						0_{5\times20} & 1_{5\times5} & 0_{5\times10} & 0_{5\times50} & 0_{10\times15} & 0_{5\times q-100}\\[3pt]
						0_{20\times20} & 0_{20\times5} & 1_{20\times10} & 0_{20\times50} &  0_{20\times15} & 0_{20\times q-100}\\[3pt]
						0_{50\times20} & 0_{50\times5} & 0_{50\times10} & 1_{50\times50} & 0_{50\times15} & 0_{50\times q-100}\\[3pt]
						0_{15\times20} & 0_{15\times5} & 0_{15\times10} & 0_{15\times50} & 1_{15\times15} & 0_{15\times q-100}\\[3pt]
						0_{p-100\times20} & 0_{p-100\times5} & 0_{p-100\times10} & 0_{p-100\times50} & 0_{p-100\times15} & 0_{p-100\times q-100} \\
					\end{pmatrix}
				\end{equation}

The population covariance matrices describing each of the ${\bf X}$ and ${\bf Y}$ random variables ($\boldsymbol{\Sigma_{XX}}$ and $\boldsymbol{\Sigma_{YY}}$) are created to establish relationships between the known canonical groups with sufficient noise (and spurious correlations) when compared to the remaining dimensions.  The underlying correlation structure for each dataset is an identity matrix except at the corresponding non-zero entries of ${\bf B}$ for which there is a correlation of 0.2.  That is, the first 10 dimensions of the ${\bf X}$ random variable have a pairwise correlations of 0.2; the next 5 dimensions of the ${\bf X}$ random variable have pairwise correlations of 0.2, etc.

Because each of the correlations between the dimensions of the ${\bf Y}$ random variable is given by a combination of $\boldsymbol{\Sigma_{YY}}$ {\em and} the constructed relationship between ${\bf X}$ and ${\bf Y}$, the variance of each ${\bf Y}$ random variable needs to be moderated to create ${\bf Y}$ random variables with specified correlations.  We set the variance of ${\bf Y}$ assuming that ${\bf Y}$ is a sum of ${\bf X}$ values as well as an error term.  See the appendix for the derivation of the $\Sigma_{YY}$ matrix.

Clean data were simulated as above according to a multivariate normal distribution.  Data with heavier tails is given using the multivariate normal set up as above with the additional modification that each multivariate normal observation is divided by the square root of $\chi^2_2$ random variable divided by its degrees of freedom.  Specifically, each of the $n$ $p$-dimensional ${\bf x}$ vectors is divided by a $\chi^2_2$ random variable divided by 2 and then used to generate the corresponding ${\bf y}$ vector.   We refer to the heavy tailed data as $t-$like data (as the original normal random deviate is neither centered at zero nor scaled to have variance one).



\subsubsection{Complete Groups \label{subsubsectionCG}}

Note that the structure of ${\bf B}$ leads to the idea of a {\em complete group}.  The notion of a complete group will be important to the assessment of the methods described in the paper.  We define a complete group to be the set of dimensions of ${\bf X}$ and of ${\bf Y}$ which are correlated.  In the simulation above, there are five complete groups given by ${\bf B}$ in equation \ref{Bmatrix}.  For example, the first complete group is represented by the dimensions $\{ (1,2,\ldots,10) \}$ in ${\bf X} \ \   \&  \ \ \{ (1,2,\ldots,20) \}$ in ${\bf Y}$ .   An incomplete group might be, for example,  $\{ (1,4, 7) \}$ in ${\bf X} \ \ \& \ \  \{ (9,15,20) \}$ in ${\bf Y}$.  The variables would be all true positives, but the overall complete relationship would not show up as having established a complete group of parameters.  The parameters of the model are the non-zero coefficients on the complete group elements only.

\subsubsection{Determining Significance \label{subsubsectionSig}}

\begin{figure}[H]
\begin{center}
\includegraphics[scale=.5,angle=0]{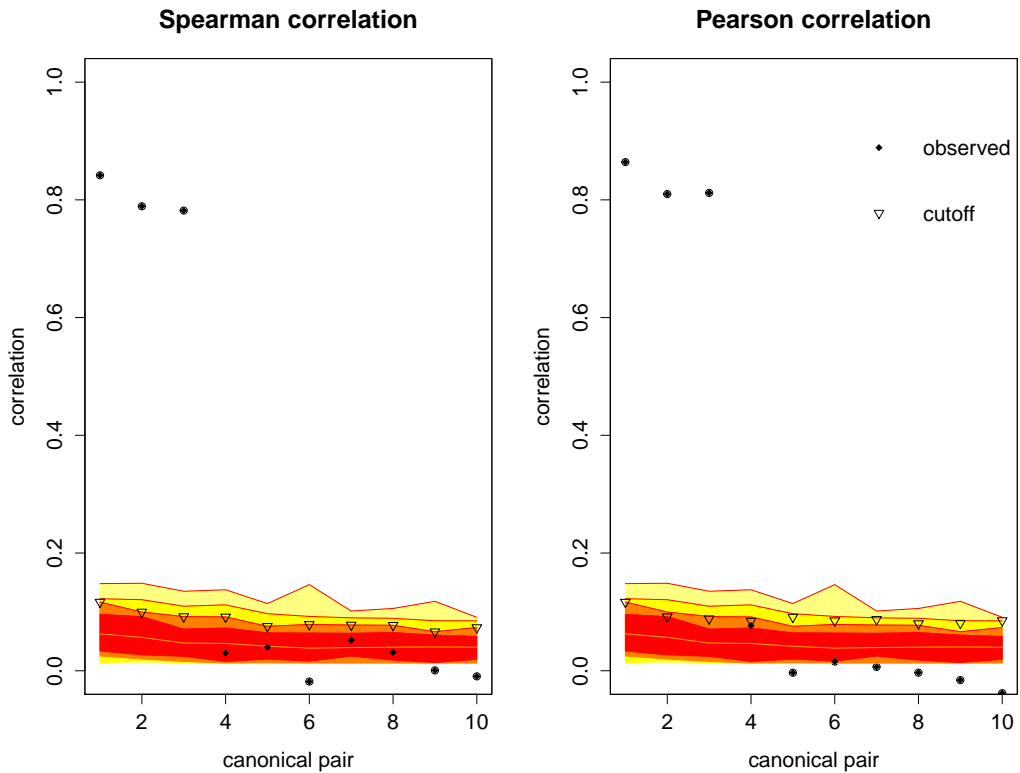}
\end{center}
\caption{\label{corFan.fig} Given one simulated dataset, both the permuted correlations (red gradient) and observed correlations (black dot) are plotted.  Additionally, the cutoff for significance is given by the 0.9 quantile of the permuted correlations (triangles) and can be seen to determine three canonical correlations (black dots) as significant.  The red gradient shows additional quantiles of the permuted distribution.}
\end{figure}

As outlined in Algorithm \ref{AlgorithmCutoff}, we use a permutation scheme to determine the cutoff values for a vector of canonical correlations (keeping in mind that they {\em decrease} across canonical pairs).  We provide a graphical representation of the algorithm to determine significance of a canonical pair.  Figure \ref{corFan.fig} plots the permuted correlations and observed correlations as a function of canonical pair for one simulated dataset.  The black dots are the observed canonical correlations, and the triangles represent the 0.9 quantile of the permuted correlations for a given canonical pair.   For the simulated dataset shown, there are three significant canonical pairs, as at the fourth canonical pair, the observed correlation (black dot) falls below the 0.9 quantile (0.9 chosen arbitrarily to be the value of the Q cutoff) of the permuted canonical correlations (triangle).

\subsubsection{Evaluation Metrics \label{SubsectionEval}}

In order to evaluate the methods described above, we compare the non-zero (i.e., non-sparse) coefficients to the original matrix used to generate the linear dependency between the random variables ${\bf X}$ and ${\bf Y}$. Recall that the response variable is generated such that for each individual $l$, ${\bf Y}_l  \sim MVN_q(\boldsymbol{\mu}_l, \boldsymbol{\Sigma_{YY}})$, where $\boldsymbol{\mu}_l = {\bf X}_l \times {\bf B}$.  The matrix ${\bf B}$ is given in equation (\ref{Bmatrix}).  We detail the evaluation metrics below, keeping in mind that each of the measurements is done for those canonical pairs whose canonical correlation is above the permutation cutoff value described in Section \ref{SectionCutoff}.  In the evaluation metrics below, we use the word {\em true} to indicate a variable which has a non-zero entry in the ${\bf B}$ matrix used to simulate the data, see equation (\ref{Bmatrix}).

\begin{itemize}

\item[NC Pair] The Number of Canonical Pairs which are significant according to the permutation test.
\item[TPR] True Positive Rate measures the total number of non-zero coefficients that are true (with double counting) divided by the sum of (total number of non-zero coefficients that are true) + (total number of empirical non-zero coefficients that are not in any complete group).  That is, the ratio of total number of coefficients that are true and non-zero divided by the total number of coefficients that have empirical non-zero coefficients.  The result is to measure the proportion of non-zero coefficients which are true.
\item[TP of CG]  True Positive of Complete Groups gives another measure of true positives.  True Positives of the Complete Groups represents the number of canonical pairs containing a complete group divided by the number of canonical pairs (NC Pair).
\item[FN Rate]  The False Negative Rate measures the number of true variables  with zero coefficients across all of the significant canonical pairs (out of a total number of true variables given in the model, e.g., see equation (\ref{Bmatrix})).
\end{itemize}

\subsection{Simulation Results \label{SubsectionResults}}

For the simulation study, we set $p=500, q=1000$ and let $n$ vary along (50,100,500,1000).  Each simulation was run 100 times; additionally, both $\lambda_u$ and $\lambda_v$ were set to range along the vector (0,0.1,0.2,0.3,0.4,0.5).  By incorporating the adjusted covariance matrix into Algorithm \ref{AlgorithmSCCA}, we are able to find the sparse loadings associated with each canonical pair.  Additionally, each canonical pair is assessed to determine whether or not it contains a complete group.  The values of the evaluation metrics above are presented below for both the clean and the $t-$like data across different values of the sample size.

\subsubsection{Number of Canonical Pairs}

The number of canonical pairs considered to be significant was determined using the permutation method (with 100 permutations) in Algorithm \ref{AlgorithmCutoff}. As mentioned above, the model was set-up to have 5 canonical pairs as given in equation (\ref{Bmatrix}). For t-like data, both MSCCA and RMSCCA tend to give more canonical pairs than the model specifies, once the sample size is sufficiently large. This is due to the complex nature of the relationships, whereas early canonical pairs may correctly find signal in the data, they may not include every in variable in the relationship (a so called complete group).  Thus, later canonical pairs may be again correctly find remaining signal from the same relationship, resulting in more estimated pairs than true pairs, though all estimates are identifying signal in the data.

\begin{figure}[H]
\begin{center}
\includegraphics[scale=.7,angle=0]{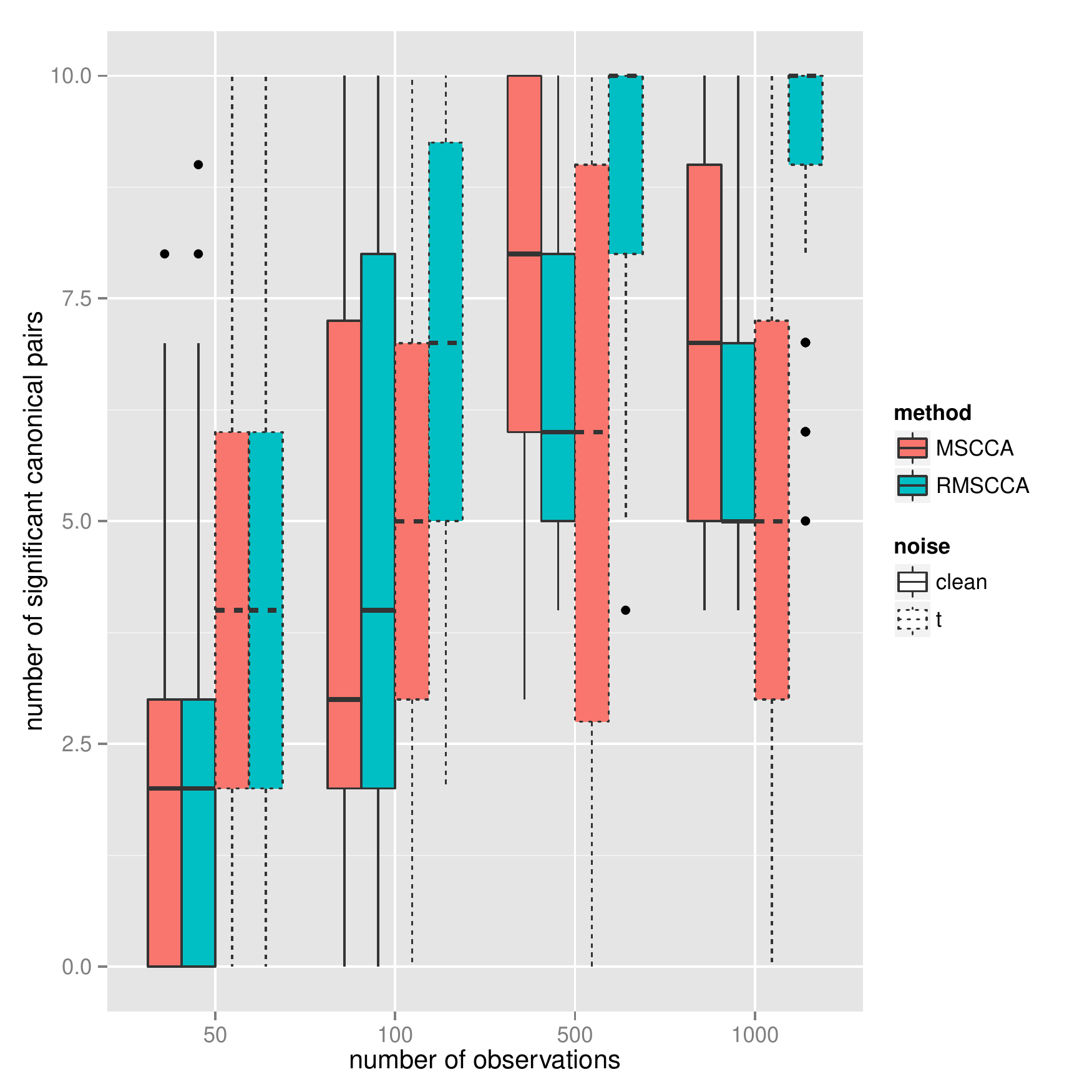}
\end{center}
\caption{\label{ncc.fig} For each of 100 simulations, the number of canonical pairs which were determined to be significant under the permutation structure is given as a function of the simulation data size.}
\end{figure}

\subsubsection{True Positive Rate}

\begin{figure}[H]
\begin{center}
\includegraphics[scale=.7,angle=0]{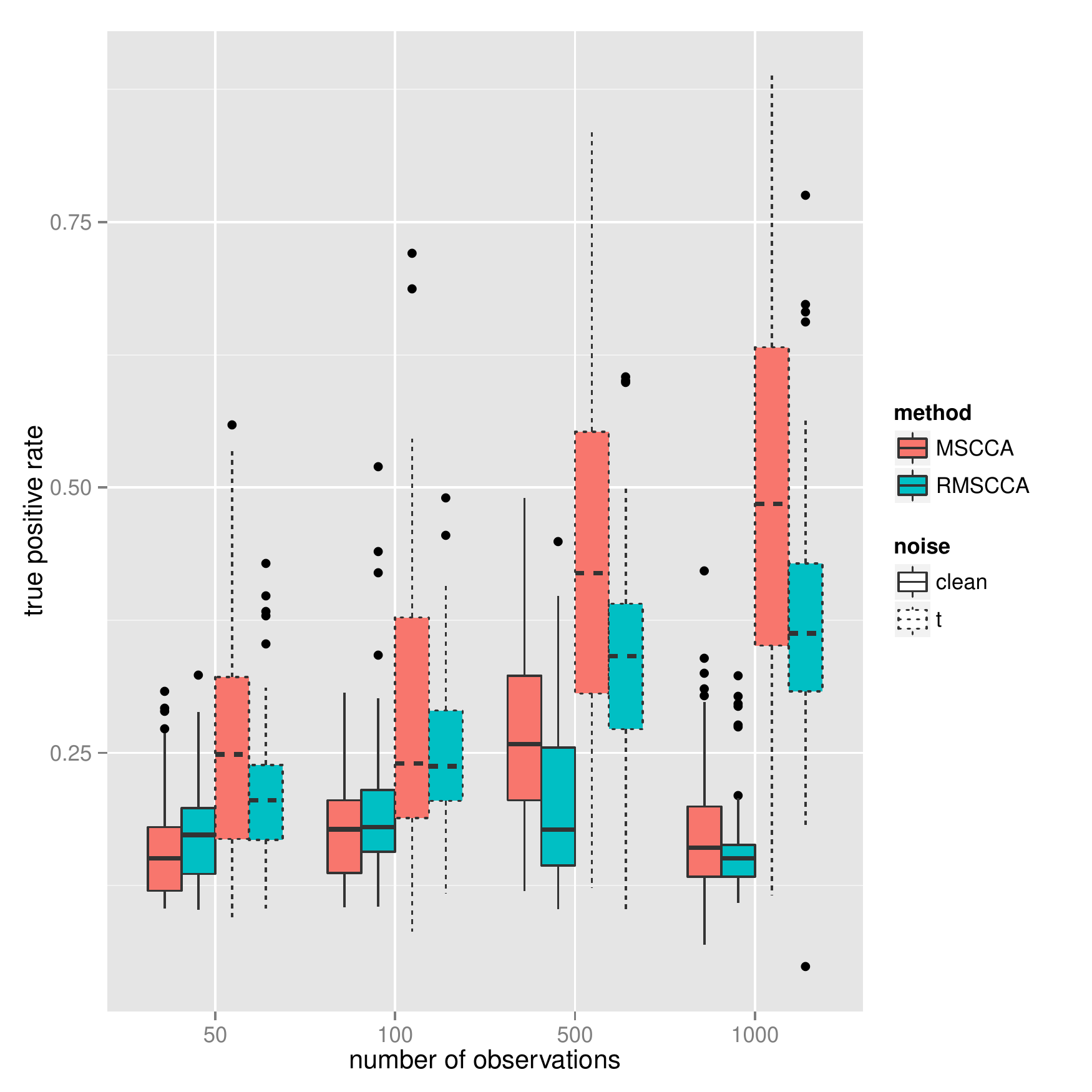}
\end{center}
\caption{\label{TPrate.fig} For each of 100 simulations, the proportion of non-zero coefficients which are true as a function of the simulation data size.}
\end{figure}

\begin{figure}[H]
\begin{center}
\includegraphics[scale=.7,angle=0]{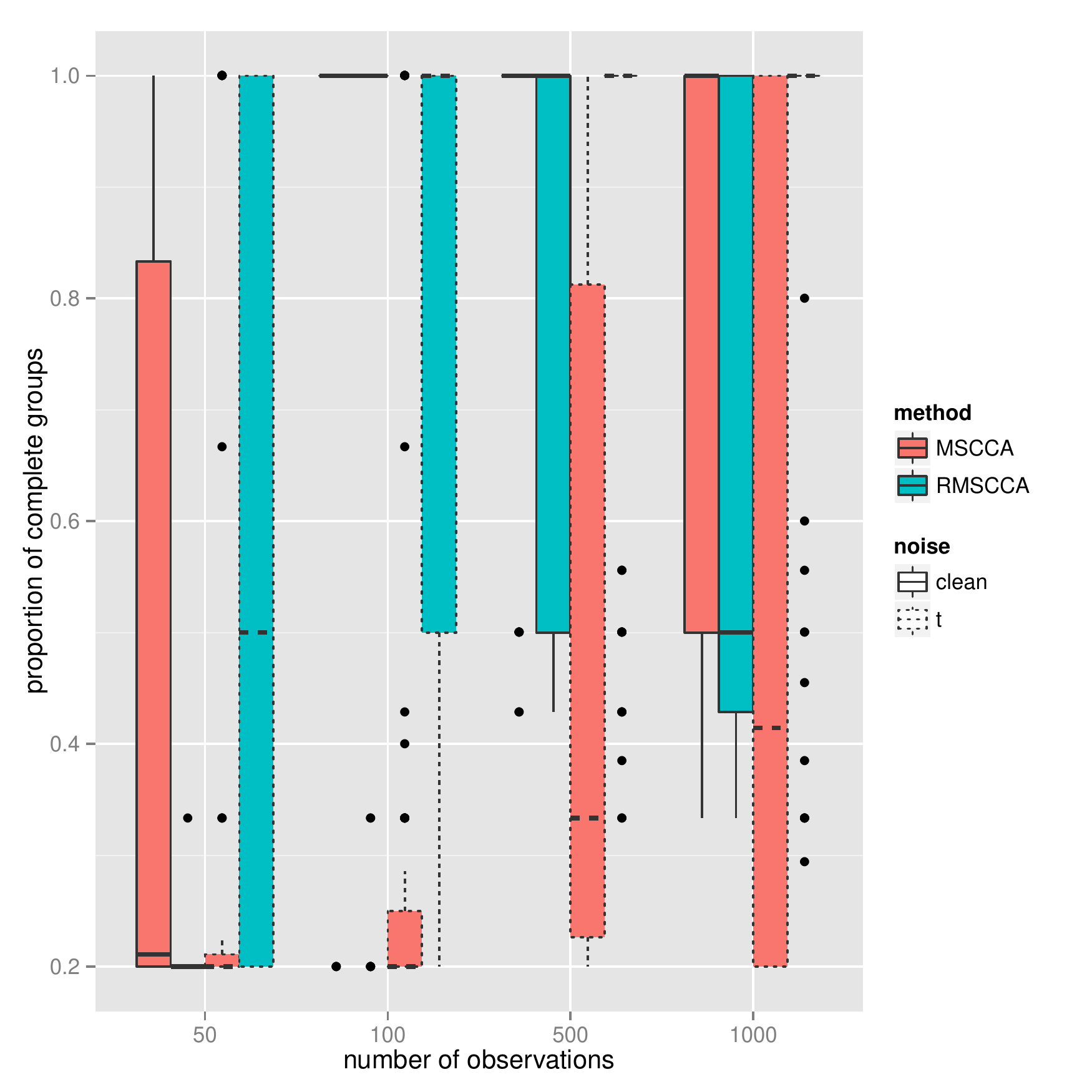}
\end{center}
\caption{\label{TPCG.fig} For each of 100 simulations, the proportion of complete groups out of the number of canonical pairs.}
\end{figure}

We measure true positives using two different metrics. The true positive rate (TPR) (see Figure \ref{TPrate.fig}) gives the proportion of non-zero coefficients across all canonical pairs.  The true positive rate of complete groups (see Figure \ref{TPCG.fig}) gives the proportion of complete groups out of the number of canonical pairs. For the TPR, we see a somewhat surprising result in that RMSCCA is lower across all sample sizes for t-like data.  This needs to be understood in terms of Figure \ref{TPCG.fig}. Whereas there are indeed a higher proportion of non-zero coefficients associated with MSCAA, this is a consequence of having overly sparse solutions.    For samples sizes of $n=100$ and higher, RMSCCA has a median complete group proportion of 1, drastically outperforming its nonresistant counterpart (whose median values, across sample sizes, never exceed 0.42).

\subsubsection{False Negative Rate}

\begin{figure}[H]
\begin{center}
\includegraphics[scale=.7,angle=0]{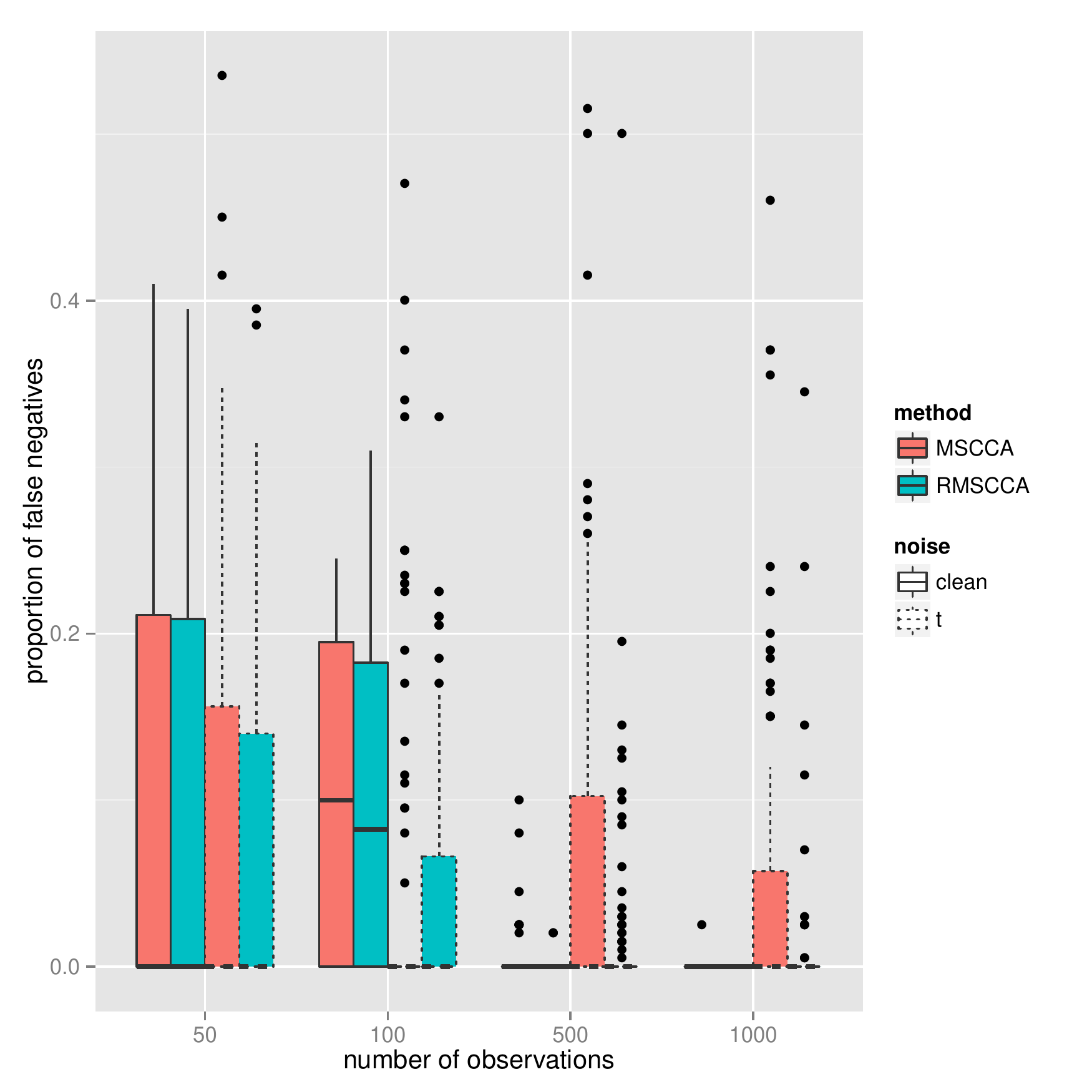}
\end{center}
\caption{\label{fn.fig} For each of 100 simulations, the false negative rate as a function of the simulation data structure. The clean data represent the first two boxes in each set, and the $t-$like data represent the second two boxes in each set.  }
\end{figure}

The False Negative Rate measures the proportion of true coefficients (see equation (\ref{Bmatrix})) which had zero coefficients for all of the significant canonical pairs (see Figure \ref{fn.fig}).  With large sample sizes, we see that only MSCCA on the $t-$like data has a substantial loss of power in determining positive coefficients across the significant canonical pairs.   Even for lower sample sizes, RMSCCA outperforms MSCCA.  (N.b., the red bar for $n=100$ with MSCCA on t-like data is missing only due to the small number of replications (100) in the simulation.  If the box plots had been made at the 0.77 quantile instead of the 0.75 quantile, the red bar would not have disappeared.)

\subsubsection{Type I errors \label{SubsubsectionTypeI}}

To confirm that the familywise error rate on null data is controlled at the 0.1 level (chosen by Q=0.9), we simulate data with no structure between $X$ and $Y$  (i.e, $B \equiv 0$).  We run the complete MSCCA and RMSCCA algorithms.  As above, we set $p=500$ and $q=1000$ letting $n$ vary on $(50,100,500,1000)$.  For each of 100 simulations, we count the number of times the observed correlations are considered significant according to the permutation scheme.  With null data, we expect to see the observed data above the cutoff $10\%$ of the time because we use a 0.9 quantile cutoff.  Table \ref{type1.tab} gives the empirical type I error rates.

\begin{table}[H]
\begin{center}
\begin{tabular}{lrrrr}
& \multicolumn{4}{c}{Sample Size, $n$}\\
& 50 & 100 & 500 &1000\\
\cline{2-5}
MSCCA & 0.06 & 0.05 & 0.10 & 0.11\\
RMSCCA & 0.04 & 0.10 & 0.09 & 0.13\\
\end{tabular}
\end{center}
\caption{\label{type1.tab} Type I error rates for data simulated as in section \ref{SubsectionSetup} with $B=0$ so as to remove the relationship between ${\bf X}$ and ${\bf Y}$.  Our method accurately controls the type I error rate at 0.1.}
\end{table}

\subsubsection{Power}

Power was calculated as the percent of simulation where at least one canonical correlation was above the permutation threshold.  Power was calculated on all of the simulations where there was signal in the data, and so the method should have given canonical pairs above the permutation threshold.  The power calculation below does not address the number of canonical pairs above the threshold.

\begin{table}[H]
\begin{center}
\begin{tabular}{lrrrr||rrrr}
& \multicolumn{8}{c}{Sample Size, $n$}\\
& \multicolumn{4}{c}{clean data} & \multicolumn{4}{c}{t-like data}\\
& 50 & 100 & 500 &1000 &50 & 100 & 500 &1000\\
\cline{2-8}
MSCCA & 0.68& 0.81 &  1 &   1 &0.84& 0.91 &0.91 &0.92\\
RMSCCA & 0.70& 0.82 &  1 &   1 & 0.87 &1.00& 1.00& 1.00\\
\end{tabular}
\end{center}
\caption{\label{pow.tab} Power calculations for data simulated with $B$ so as to construct the relationships between ${\bf X}$ and ${\bf Y}$ described in section \ref{SubsectionSetup}.  The power is seen to be higher for the resistant method across the board.}
\end{table}

\subsection{Real Data \label{SubsectionData}}
Next, we applied RMSCCA to a real biological dataset to compare our method to that of \citet{Witten}.

We analyzed the \citet{Chin06} copy number abnormality (CNA) and mRNA expression data available in the PMA package in R \citep{PMA} in order to facilitate these comparisons.

\citet{Chin06} measured mRNA expression of p=19672 genes on Affymetrix U133A microarrays and measured q=2149 CNAs on Bacterial Artificial Chromosome (BAC) Comparative Genomic Hybridization (CGH) arrays in aggressively treated early-stage breast tumors obtained from $n$=89 subjects. Although log-transformed CNA and microarray data are often {\em assumed} to follow a normal distribution, both data types are more accurately described by a heavy tailed distribution \citep{Hardin09,Roy13}, and failure to account for this distribution can lead to spurious associations. Notably, many types of biological data, like genotype data, methylation data, and clinical outcomes, deviate from the assumption of normality. While non-normal data should not be accommodated by a classical SCCA algorithm, our employment of resistant estimation makes our method suitable.

Here, we applied RMSCCA to the copy number and mRNA expression data from each of the 23 chromosomes separately. Unlike \citet{Witten} who set their tuning parameters to achieve a sparse solution including only $\approx$25 coefficients, we used cross validation as outlined in Algorithm \ref{AlgorithmCV} to set our tuning parameters. Most importantly, our RMSCCA algorithm allowed us to consider multiple canonical pairs per each chromosome and to assess the significance of these multiple canonical pairs using our permutation test as outlined in Algorithm \ref{AlgorithmCutoff} (with the maximum number of canonical pairs set to 10).  The algorithm took between a few hours up to 45 hours to run for a given chromosome.    The analysis was performed on a computer with two eight core AMD Opteron 6276 processors running at 1.4 GHz.  The analysis can also be parallelized for a reduction in computational time.


We tested the significance for each of the top ten canonical pairs using all 23 chromosomes (compared individually).  We see that the majority of the chromosomes have 10 significant canonical pairs, but not all of them.  Indeed, some of the chromosomes have no significant canonical pairs, see Table \ref{brsig.tab}.  Though different from the analysis of \citet{Witten} for the reasons given above, our analysis is consistent with theirs in the sense that much of the signal within chromosomes is significant.

\begin{table}[ht]
\centering
\begin{tabular}{rrr}
  \hline
Chromosome & \# Signif (RMSCCA) & \# Signif (MSCCA) \\
  \hline
1 & 10 & 10 \\
  2 & 2 & 4 \\
  3 & 3 & 10 \\
  4 & 10 & 7 \\
  5 & 10 & 10 \\
  6 & 10 & 10 \\
  7 & 10 & 10 \\
  8 & 10 & 10 \\
  9 & 10 & 10 \\
  10 & 10 & 1 \\
  11 & 10 & 10 \\
  12 & 10 & 10 \\
  13 & 10 & 10 \\
  14 & 10 & 10 \\
  15 & 10 & 10 \\
  16 & 10 & 10 \\
  17 & 0 & 8 \\
  18 & 10 & 10 \\
  19 & 10 & 4 \\
  20 & 0 & 2 \\
  21 & 10 & 10 \\
  22 & 10 & 10 \\
  23 & 0 & 0 \\
   \hline
\end{tabular}
\caption{\label{brsig.tab}  Using the breast cancer data of \cite{Chin06}, for each chromosome, the number of significant canonical pairs.}
\end{table}

\begin{figure}[ht]
\begin{center}
\includegraphics[scale=.4,angle=0]{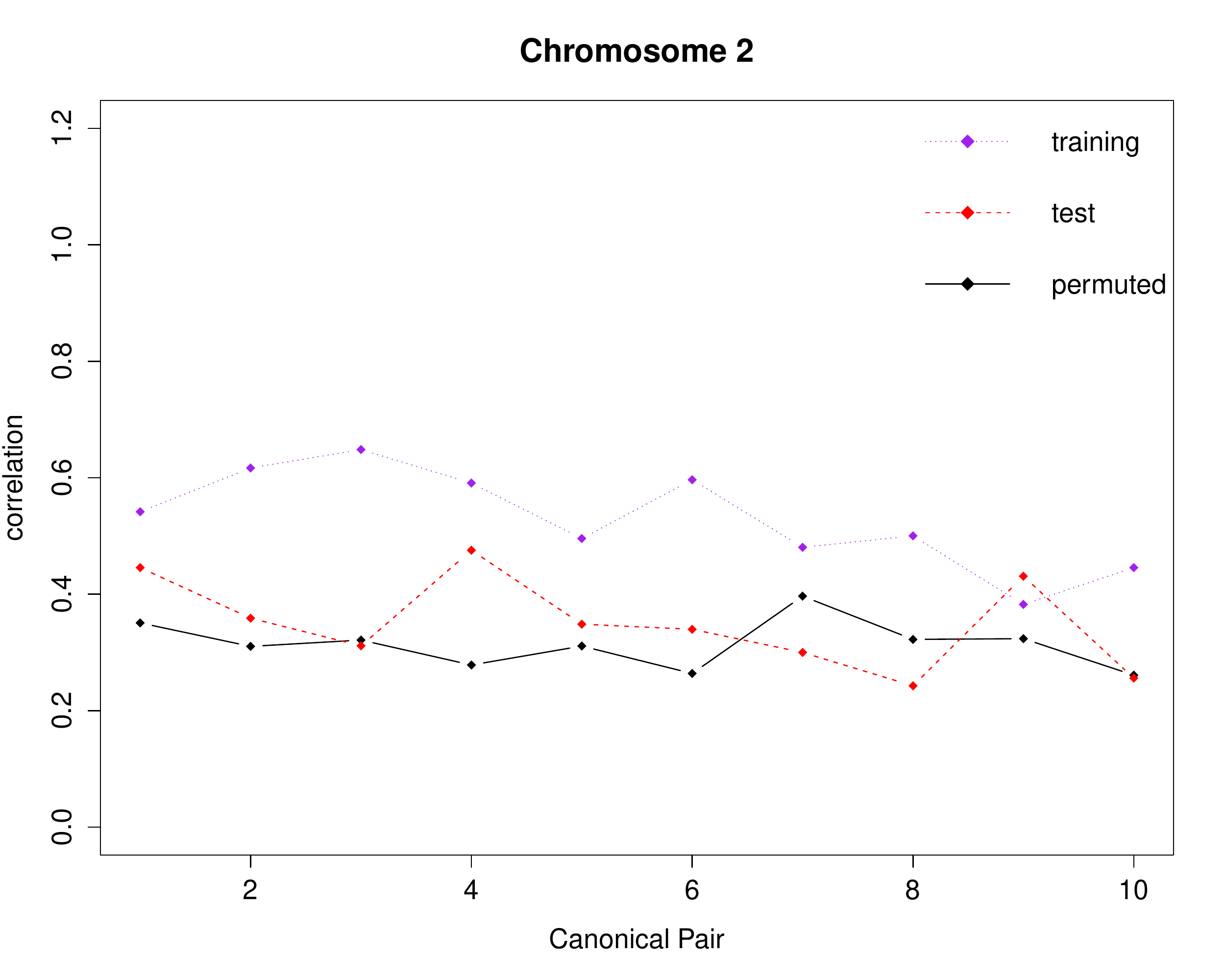}
\end{center}
\caption{\label{corChrom2.fig} Using the breast cancer dataset we calculate the canonical correlation from chromosome 2 for the first 10 canonical pairs in three ways: 1. purple: canonical correlations for the full dataset based on coefficients calculated using the full dataset; 2. red: canonical correlations averaged over the cross validation procedure, coefficients from the training data, correlations on the test data; 3. the 0.9 quantile of the canonical correlations from the cross validation procedure over the {\em permuted} dataset.}
\end{figure}

A closer look at chromosome 2 (using RMSCCA) shows that except for the first two canonical pairs, the test data is not significantly different from the permuted data, see figure \ref{corChrom2.fig}.  It is important to point out the purple training data points such that they suffer from the curse of dimensionality.  Comparing the permuted data correlations to the training data correlations would not have been an accurate comparison due to the huge dimensionality and over-fitting that happens through the canonical correlation estimation process.

\begin{figure}[ht]
\begin{center}
\includegraphics[scale=.8,angle=0]{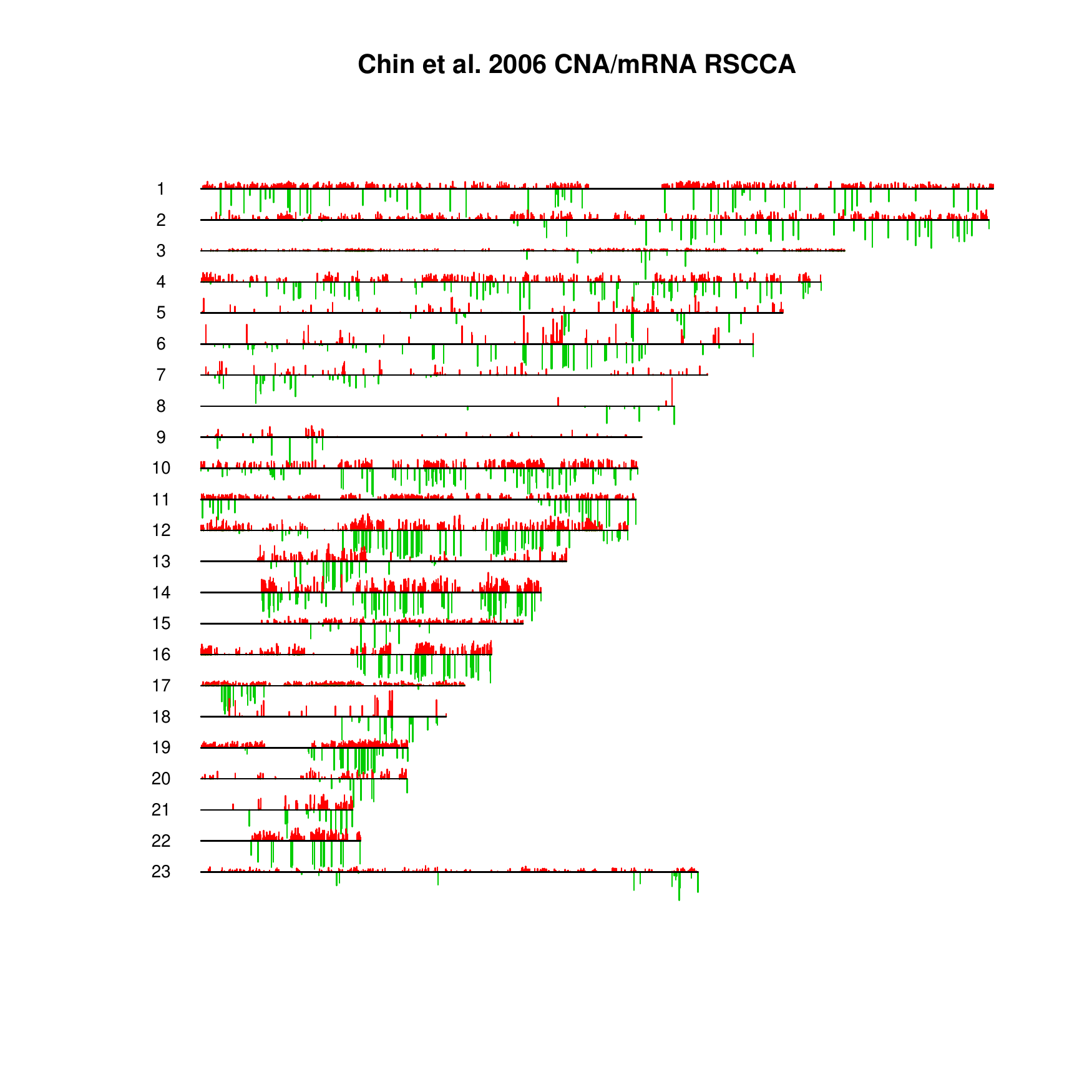}
\end{center}
\caption{\label{brcoef.fig} Using the breast cancer dataset of \citet{Chin06}, we provide the canonical coefficients for the first canonical pair across each of the 23 chromosomes.  The red ticks (ticks above the horizontal line) represent the canonical coefficients associated with the mRNA data and the green ticks (ticks below the horizontal line) represent the canonical coefficients associate with the CNA data.  For each tick, its location is given by the placement along the chromosome and the length of the tick is proportional to the magnitude of the canonical coefficient value.}
\end{figure}

The canonical coefficients for the first canonical pairs (using RMSCCA) across each of 23 chromosomes is given in Figure \ref{brcoef.fig}.  The red ticks represent the mRNA coefficients (both chromosomal location and magnitude of coefficient) and the green ticks represent the CNA coefficients (both chromosomal location and magnitude of coefficient).  As in the analysis by \citet{Witten}, we see that the strong correlations (i.e., high canonical correlations) are given by comparing mRNA and CNA variables which are located at the same points along the given chromosome.

\section{Conclusion \label{SectionConclusion}}

Canonical correlation analysis gives linear relationships between variables from two distinct datasets.  We have extended previous work on sparse canonical correlation to allow for both multiple canonical pairs and for resistant analysis in the setting where $n << p, q$.  Our work shows that the method is able to find the simulated structure both in terms of number of canonical pairs and in terms of complete groups. Like MSCCA, RMSCCA still gives a large number of coefficients, but the true variables are typically returned.  We see also that with heavy tailed data, it is better to use resistant correlation to avoid any leverage points.  The analysis of a real dataset gives consistent results to that of \citet{Witten} in terms of both significance as well as connecting the chromosomal locations of the mRNA and CNA measurements.

\section*{Appendix}

Consider the case of the first dimension of ${\bf Y}$, $Y_{1}$,  which is centered at the first $p_1$ dimensions of the random variable ${\bf X}$.  Because the majority of the correlation between the dimensions of the random variable ${\bf Y}$ values comes from their dependence on the random variable ${\bf X}$, let $\Sigma_{YY}$ be a diagonal matrix.  In contrast, $\boldsymbol{\Sigma_{XX}}$ is made up of $\rho (=0.2)$ at the appropriate off diagonal elements and 1 on the diagonal.

Below is the derivation for the first diagonal entry of $\Sigma_{YY}$, $\sigma_{YY,11}$.  The goal is to find $\sigma_{YY,11}$ such that cor($y_{l1}, y_{l2}$) = $\rho$.

\begin{eqnarray*}
{\bf Y}_l  &\sim& MVN_q(\mu_l, \Sigma_{YY}), \mbox{ where } \boldsymbol{\mu}_l = {\bf X}_l \times B, \ \ l=1, \ldots, n\\
{\bf Y}_l &=& {\bf X}_l \times B + \epsilon_l, \mbox{ where } \epsilon_l \sim MVN_q(0, \Sigma_{YY}), \ \ l=1, \ldots, n\\
Y_{l1} &=& \sum_{i=1}^{p_1} {X_{li}} + \epsilon_{l1}, \mbox{ where } \epsilon_{l1} \stackrel{iid}{\sim} N(0, \sigma_{YY,11})\\
&&\\
Var(Y_{l1}) &=& Var\bigg(\sum_{i=1}^{p_1} {X_{li}} + \epsilon_{l1}\bigg)\\
&=& p_1 \sigma_{XX,11} + (p_1^2 - p_1)\sigma_{XX,12} + Var(\epsilon_{l1})  \ \ \ \mbox{WLOG}\\
Var(Y_{l1}) &=& p_1  + (p_1^2 - p_1)\rho + \sigma_{YY,11}\\
&&\\
Cov(Y_{l1}, Y_{l2}) &=& Cov\bigg(\sum_{i=1}^{p_1} {X_{li}} + \epsilon_{l1}, \sum_{i=1}^{p_1} {X_{li}} + \epsilon_{l2}\bigg)\\
&=& p_1\sigma_{XX,11} + p_1(p_1-1) \sigma_{XX,12} + cov(\epsilon_{l1}, \epsilon_{l2})   \ \ \ \mbox{WLOG}\\
 &=& p_1 + p_1(p_1-1) \rho\\
&&\\
Cor(Y_{l1}, Y_{l2}) &=& \frac{p_1 + (p_1^2 - p_1) \rho}{p_1  + (p_1^2 - p_1)\rho + \sigma_{YY,11}} = \rho\\
\sigma_{YY,11} &=& \bigg( \frac{1}{\rho} -1 \bigg) (p_1 + (p_1^2 - p_1)\rho)
\end{eqnarray*}

By increasing the variance for each of the simulated ${\bf Y}$ variables involved in the true linear relationships, we create correlations of $\rho$ (=0.2 in our simulations) between the ${\bf Y}$ variables in a group.   The cross-covariance matrix between ${\bf X}$ and ${\bf Y}$ ($\boldsymbol{\Sigma_{XY}}$) is not pre-specified, but rather it is given by the relationship between $\boldsymbol{\Sigma_{XX}}$, $\boldsymbol{\Sigma_{YY}}$, and ${\bf B}$.

\renewcommand{\baselinestretch}{1} \large \normalsize

\bibliographystyle{abbrvnat}
\bibliography{rsmccaRef}
\end{document}